\documentclass[sigconf, nonacm]{acmart}

\newcommand\vldbdoi{10.14778/3828612.3828629}
\newcommand\vldbpages{2755 - 2762}
\newcommand\vldbvolume{19}
\newcommand\vldbissue{10}
\newcommand\vldbyear{2026}
\newcommand\vldbauthors{Matteo Brucato, Fjodor Kholodkov, Soren Little, Jakob Mayer, and Duc Nguyen}
\newcommand\vldbtitle{\systemname: Prescriptive Analytics in the Data Stack}
\newcommand\vldbavailabilityurl{https://github.com/osm-data/decisionhouse-analysis}
\newcommand\vldbpagestyle{empty} %

\usepackage{etoolbox}
\usepackage{booktabs}
\usepackage{xcolor}
\usepackage{colortbl}
\usepackage{listings}
\usepackage{amsmath}
\usepackage{graphicx}
\usepackage{tikz}
\usetikzlibrary{positioning, arrows.meta, calc, decorations.pathreplacing, shapes.geometric}
\usepackage{microtype}
\usepackage{multirow}
\usepackage{balance} %
\usepackage{enumitem} %
\usepackage[capitalise,noabbrev]{cleveref} %
\setlist{leftmargin=*, nosep, topsep=2pt}

\newcommand{\systemname}{Decisionhouse}           %
\newcommand{\productname}{DH-0}
\newcommand{\querylang}{D\lowercase{e}QL}                     %
\newcommand{\systemnameplural}{Decisionhouses}    %
\newcommand{\optimizername}{DeOP}                 %
\newcommand{\numrequirements}{eight}              %
\newcommand{\Numrequirements}{Eight}              %
\newcommand{\reqname}[1]{{\setlength{\fboxsep}{1.5pt}\colorbox{gray!15}{#1}}}

\usepackage{needspace}

\newif\ifextended
\extendedfalse    %

\newif\ifappendix
\ifextended\appendixtrue\else\appendixfalse\fi

\newif\ifarxiv
\arxivtrue
\ifextended\arxivtrue\fi

\newif\ifpaperonly
\paperonlytrue   %

\usepackage{pifont}
\usepackage{wasysym}
\newcommand{\cmarkv}{\textcolor{green!70!black}{\ding{51}}}                          %
\newcommand{\pmarkv}{\textcolor{orange!80!black}{$\circ$}}                           %
\newcommand{\xmarkv}{\textcolor{red!80!black}{\ding{55}}}                            %

\usepackage{tikz}
\usetikzlibrary{positioning, arrows.meta, calc, fit, backgrounds, decorations.pathreplacing}

\AtEndEnvironment{figure}{\vspace{-3pt}}
\AtEndEnvironment{figure*}{\vspace{-3pt}}
\AtEndEnvironment{table}{\vspace{-3pt}}
\AtEndEnvironment{table*}{\vspace{-3pt}}

\definecolor{kwcolor}{RGB}{0,80,160}      %
\definecolor{strcolor}{RGB}{163,21,21}    %
\definecolor{numcolor}{RGB}{128,0,128}    %
\definecolor{colcolor}{RGB}{70,70,110}    %
\definecolor{relcolor}{RGB}{0,110,70}     %
\definecolor{concolor}{RGB}{140,80,140}   %
\definecolor{catalogcolor}{RGB}{180,80,0} %
\definecolor{deccolor}{RGB}{180,85,0}     %

\definecolor{R2color}{RGB}{0,80,160}    %
\definecolor{R3color}{RGB}{0,120,60}    %
\definecolor{R4color}{RGB}{200,100,0}   %
\definecolor{Rgencolor}{RGB}{120,40,140} %
\colorlet{R2COLOR}{R2color}
\colorlet{R3COLOR}{R3color}
\colorlet{R4COLOR}{R4color}
\colorlet{RGENCOLOR}{Rgencolor}
\newif\ifrevision
\revisiontrue
\newcommand{\hlRtwo}[1]{\ifrevision\textcolor{R2color}{#1}\else#1\fi}
\newcommand{\hlRthree}[1]{\ifrevision\textcolor{R3color}{#1}\else#1\fi}
\newcommand{\hlRfour}[1]{\ifrevision\textcolor{R4color}{#1}\else#1\fi}
\newcommand{\hlgen}[1]{\ifrevision\textcolor{Rgencolor}{#1}\else#1\fi}
\newif\ifarxivdiff\arxivdifffalse  %
\definecolor{ArXivNewColor}{rgb}{0.75,0.10,0.10}
\newcommand{\hlnew}[1]{\ifarxivdiff\textcolor{ArXivNewColor}{#1}\else#1\fi}

\newcommand{\revref}[1]{\hyperref[rev:#1]{\textbf{#1}}}

\newif\ifwithletter
\withlettertrue

\ifpaperonly
  \withletterfalse
  \revisionfalse
\fi
\ifwithletter
\usepackage{tcolorbox}
\newtcolorbox{reviewquote}[1][]{
  colback=gray!8, colframe=gray!40,
  boxrule=0.4pt, arc=2pt, left=4pt, right=4pt, top=2pt, bottom=2pt,
  before skip=10pt, after skip=2pt,
  fontupper=\small\itshape, #1
}
\newtcolorbox{metaquote}[1][]{
  colback=gray!8, colframe=gray!40,
  boxrule=0.4pt, arc=2pt, left=4pt, right=4pt, top=2pt, bottom=2pt,
  before skip=10pt, after skip=2pt,
  fontupper=\small\itshape, #1
}
\newtcolorbox{rtwoquote}[1][]{
  colback=R2color!5, colframe=R2color!60,
  boxrule=0.5pt, arc=2pt, left=4pt, right=4pt, top=2pt, bottom=2pt,
  before skip=10pt, after skip=2pt,
  fontupper=\small\itshape, #1
}
\newtcolorbox{rthreequote}[1][]{
  colback=R3color!5, colframe=R3color!60,
  boxrule=0.5pt, arc=2pt, left=4pt, right=4pt, top=2pt, bottom=2pt,
  before skip=10pt, after skip=2pt,
  fontupper=\small\itshape, #1
}
\newtcolorbox{rfourquote}[1][]{
  colback=R4color!5, colframe=R4color!60,
  boxrule=0.5pt, arc=2pt, left=4pt, right=4pt, top=2pt, bottom=2pt,
  before skip=10pt, after skip=2pt,
  fontupper=\small\itshape, #1
}

\fi

\newcommand{\kw}[1]{\texttt{#1}}  %
\newcommand{\kwSELECT}{\kw{SELECT}}
\newcommand{\kwPREDICT}{\kw{PREDICT}}
\newcommand{\kwDECIDE}{\kw{DECIDE}}
\newcommand{\kwSELECTION}{\kw{SELECTION}}

\newcommand{\kwCREATECANDIDATES}{\kw{CREATE CANDIDATES}}

\newcommand{\kwCREATEABSTRACTION}{\kw{CREATE ABSTRACTION}}
\newcommand{\kwDECISIONKEY}{\kw{DECISION KEY}}
\newcommand{\kwDECISIONCOLUMNS}{\kw{DECISION COLUMNS}}
\newcommand{\kwBY}{\kw{BY}}

\lstdefinelanguage{DeQL}[]{SQL}{
  morekeywords=[2]{CANDIDATES, ABSTRACTION, DECIDE, SELECTION,
                   KEY, DECISION, INTO, SUBJECT, TO, CONSTRAINT, MINIMIZE, MAXIMIZE,
                   CONTINUOUS, BINARY, USING, ATTRIBUTES, ON, BOUNDS, INTEGER, VARIABLES, WITH, COLUMNS,
                   FILTER, DISJUNCTIVE, CUMULATIVE, OVERLAPS, TIMESTAMP, INTERVAL,
                   REPEAT, SUCH, THAT, WITHIN, TIMEOUT,
                   STOCHASTIC, PROBABILITY, EXPECTED},
  morekeywords=[3]{wh_id, store_id, product_id, route_id, worker_id, shift_id, job_id, pool_id, workload_id, latency_ms,
                   item_id, recipe_id, sku, region, category, from_wh, to_store, location,
                   cost, demand, price, weight, value, profit, rating, qty, id,
                   capacity, time_hrs, max_qty, max_budget, budget_alloc,
                   skill_level, required_skill, processing_time, deadline, priority, status,
                   kcal, fat, gluten, stock, margin, in_stock, available,
                   perishable, on_sale, channel_id, expected_roi, shipping_cost, warehouse,
                   active_in_last_quarter, employee_id, duration, departure_time, arrival_time,
                   slot_id, slot_date, start_hour, flight_date, holiday_date,
                   batch_id, one_batch, batch_cap, item_deadline, sequence,
                   gpu_type, memory_gb, min_memory_gb, qos_class, mem_per_unit,
                   demand_mean, demand_std, max_servings},
  morekeywords=[4]{warehouses, stores, routes, products, items, orders, forecasts, shipping, gpu_pools, gpu_assignments, allocation_plan, assignment_clusters, assignments, workloads,
                   Inventory, Recipes, Products, Items, Workers, Shifts, Catalog, ProductionJobs,
                   MarketingChannels, Crew, Flights, Holidays, TimeSlots,
                   w, r, s, p, f, I, R, P,
                   shipping_routes, optimal_shipments, active_routes, shipping_plan, replenishment_items, restock_plan,
                   gluten_free_recipes, meal_plan, worker_shifts, schedule, jobs, production_plan,
                   product_clusters, route_clusters, cart, order, selection, result, potential_routes, channels,
                   spending_plan, plan, training_tasks, training_schedule, training_options,
                   item_batch_assignments, batch_timing, batches, a, t},
  morekeywords=[5]{supply_limit, meet_demand, time_limit, total_budget, per_channel, min_active, pool_capacity,
                   budget, capacity, one_shift, one_worker, deadline, no_overlap, count, calories,
                   total_weight, per_category, perishable_weight, no_flight_conflict, no_sunday,
                   no_holiday, no_conflict, one_per_employee, link,
                   sla_reserve, mem_capacity, min_recipes, max_recipes},
  morekeywords=[6]{active, quantity, is_selected, selected, servings, assigned, start_time, machine},
  morekeywords={BY, TRUE, FALSE, SET},
  sensitive=false,
}

\lstdefinestyle{deql}{
  language=DeQL,
  basicstyle=\ttfamily\footnotesize\linespread{0.9}\selectfont,
  keywordstyle=\color{kwcolor}\bfseries,          %
  keywordstyle=[2]\color{kwcolor}\bfseries,       %
  keywordstyle=[3]\color{colcolor},               %
  keywordstyle=[4]\color{relcolor},               %
  keywordstyle=[5]\color{concolor}, %
  keywordstyle=[6]\color{deccolor},               %
  commentstyle=\color{gray}\itshape,
  stringstyle=\color{strcolor},                   %
  numberstyle=\color{numcolor},                   %
  literate=*{0}{{{\color{numcolor}0}}}{1}         %
            {1}{{{\color{numcolor}1}}}{1}
            {2}{{{\color{numcolor}2}}}{1}
            {3}{{{\color{numcolor}3}}}{1}
            {4}{{{\color{numcolor}4}}}{1}
            {5}{{{\color{numcolor}5}}}{1}
            {6}{{{\color{numcolor}6}}}{1}
            {7}{{{\color{numcolor}7}}}{1}
            {8}{{{\color{numcolor}8}}}{1}
            {9}{{{\color{numcolor}9}}}{1},
  breaklines=true,
  showstringspaces=false,
  columns=fullflexible,
  keepspaces=true,
  basewidth={0.5em,0.4em},
  xleftmargin=3pt,
  xrightmargin=3pt,
  frame=single,
  frameround=tttt,
  framerule=0.4pt,
  rulecolor=\color{kwcolor!20},
  backgroundcolor=\color{kwcolor!3},
  aboveskip=4pt,
  belowskip=2pt,
  tabsize=2,
}

\begin{document}

\title[Decisionhouse]{\systemname: Prescriptive Analytics in the Data Stack}
\ifextended\subtitle{Extended Version}\fi

\author{Matteo Brucato\quad Fjodor Kholodkov\quad Soren Little\quad Jakob Mayer\quad Duc Nguyen}
\affiliation{%
  \institution{OSM Data}
  \institution{\{matteo, fjodor, soren, jakob, duc\}@osm-data.com}
}

\begin{abstract}
Data platforms have evolved by making data-intensive workloads native: SQL and query optimizers eliminated bespoke data-retrieval programs; Lakehouses added first-class support for ML training and serving over the same data. Prescriptive analytics (computing optimal actions subject to constraints over data) is equally data-intensive, yet remains outside the platform: every optimization problem requires a hand-built pipeline from data extraction to solver invocation, rebuilt from scratch whenever the data or the requirements change. We propose \emph{\systemnameplural{}}, a new class of data infrastructure that makes prescriptive analytics \emph{native}. A \systemname{} provides (i)~\querylang{} (Decision Query Language), a declarative SQL extension where users express decision problems over relational data; (ii)~automatic formulation selection that exploits query and data semantics to pick the right problem class and solver---a choice that can change a query's complexity class from NP-hard to polynomial; and (iii)~end-to-end integration of optimization into the data platform, from query parsing through solver execution. \systemnameplural{} can help address several challenges that have kept optimization outside data platforms, including pipeline brittleness, formulation expertise, structural blindness, and scalability cliffs, and make decision-making as accessible as querying data.

\ifarxiv\ifdefempty{\vldbavailabilityurl}{}{%
\par\smallskip\noindent{\small The source code, data, and/or other artifacts have been made available at \url{\vldbavailabilityurl}.\par}%
}\fi
\end{abstract}

\maketitle

\ifarxiv
\pagestyle{plain}
\enlargethispage{-\baselineskip}
\else
\pagestyle{\vldbpagestyle}
\begingroup\small\noindent\raggedright\textbf{PVLDB Reference Format:}\\
\vldbauthors. \vldbtitle. PVLDB, \vldbvolume(\vldbissue): \vldbpages, \vldbyear.\\
\href{https://doi.org/\vldbdoi}{doi:\vldbdoi}
\endgroup
\begingroup
\renewcommand\thefootnote{}\footnote{\noindent
This work is licensed under the Creative Commons BY-NC-ND 4.0 International License. Visit \url{https://creativecommons.org/licenses/by-nc-nd/4.0/} to view a copy of this license. For any use beyond those covered by this license, obtain permission by emailing \href{mailto:info@vldb.org}{info@vldb.org}. Copyright is held by the owner/author(s). Publication rights licensed to the VLDB Endowment. \\
\raggedright Proceedings of the VLDB Endowment, Vol. \vldbvolume, No. \vldbissue\ %
ISSN 2150-8097. \\
\href{https://doi.org/\vldbdoi}{doi:\vldbdoi} \\
}\addtocounter{footnote}{-1}\endgroup

\ifdefempty{\vldbavailabilityurl}{}{
\vspace{.3cm}
\begingroup\small\noindent\raggedright\textbf{PVLDB Artifact Availability:}\\
The source code, data, and/or other artifacts have been made available at \url{\vldbavailabilityurl}.
\endgroup
}
\fi

\section{Introduction}
\label{sec:introduction}

\begin{figure*}[t]
\centering
\resizebox{\textwidth}{!}{%
\begin{tikzpicture}[
    >=Stealth,
    every node/.style={font=\sffamily},
    lakebox/.style={draw, thick, rounded corners=3pt,
        minimum height=0.6cm, font=\sffamily\bfseries},
    abox/.style={draw, thick, rounded corners=2pt,
        minimum width=2.2cm, minimum height=0.5cm, align=center, font=\sffamily\small, inner sep=2pt},
    pbox/.style={draw, thick, rounded corners=2pt, fill=gray!20,
        minimum width=2cm, minimum height=0.35cm, align=center, font=\sffamily\footnotesize, inner sep=2pt},
    ghost/.style={draw, gray!60, thick, rounded corners=2pt, fill=white,
        minimum width=2cm, minimum height=0.35cm, align=center,
        font=\sffamily\footnotesize, text=gray!45, inner sep=2pt},
    ghost2/.style={draw, gray!80, thick, rounded corners=2pt, fill=white,
        minimum width=2cm, minimum height=0.35cm, align=center,
        font=\sffamily\footnotesize, text=gray!70, inner sep=2pt},
    arr/.style={->, thick},
]

\node[font=\sffamily\bfseries\large] at (3.75, -0.6) {(a) Today};

\node[lakebox, minimum width=12cm] (lake1) at (3.75, 0) {Data Lakehouse};

\node[abox] (desc) at (-1, 3.65) {Descriptive Analytics};
\node[abox] (pred) at (1.8, 3.65) {Predictive Analytics};
\node[abox, minimum width=5.9cm] (presc) at (6.8, 3.65) {Prescriptive Analytics};

\draw[arr] ([yshift=0pt]lake1.north -| desc) -- (desc.south);
\draw[arr] ([yshift=0pt]lake1.north -| pred) -- (pred.south);

\node[font=\sffamily\small\itshape, anchor=south, fill=white] at (0.4, 0.55) {native};

\node[ghost2] (t3) at (8.2, 0.9) {Extract Data};
\node[ghost2] (f3) at (8.2, 1.55) {Formulate};
\node[ghost2] (s3) at (8.2, 2.23) {Solve};
\node[ghost2] (r3) at (8.2, 2.9) {Extract Results};
\draw[->, gray!75, thick] (t3) -- (f3);
\draw[->, gray!75, thick] (f3) -- (s3);
\draw[->, gray!75, thick] (s3) -- (r3);

\draw[->, gray!75, thick] ([xshift=4.45cm]lake1.north) -- (t3.south);
\draw[->, gray!75, thick] (r3.north) -- (r3.north |- presc.south);
\draw[->, gray!75, thick] (r3.east) -- (9.5, 2.9) -- (9.5, 0.3);

\node[font=\Large, gray!75] at (6.8, 0.9) {\textbf{\dots}};
\node[font=\Large, gray!75] at (6.8, 1.55) {\textbf{\dots}};
\node[font=\Large, gray!75] at (6.8, 2.23) {\textbf{\dots}};
\node[font=\Large, gray!75] at (6.8, 2.9) {\textbf{\dots}};

\node[pbox] (t1) at (5.4, 0.9) {Extract Data};
\node[pbox] (f1) at (5.4, 1.55) {Formulate};
\node[pbox] (s1) at (5.4, 2.23) {Solve};
\node[pbox] (r1) at (5.4, 2.9) {Extract Results};

\draw[arr] (t1) -- (f1);
\draw[arr] (f1) -- (s1);
\draw[arr] (s1) -- (r1);

\draw[arr] ([xshift=1.65cm]lake1.north) -- (t1.south);

\draw[arr] (r1.north) -- (r1.north |- presc.south);

\draw[arr] (r1.west) -- (4.1, 2.9) -- (4.1, 0.3);

\draw[decorate, decoration={brace, amplitude=5pt}, thick]
    (10.0, 3.1) -- (10.0, 0.65)
    node[midway, right=5pt, font=\sffamily\small, align=center, fill=white, inner sep=3pt]
    {built from scratch\\per decision problem\\(\emph{decision debt})};

\node[font=\sffamily\bfseries\large] at (18.5, -0.6) {(b) With a \textsc{Decisionhouse}};

\node[draw, thick, rounded corners=3pt,
    minimum width=10.5cm, minimum height=2.2cm] (stack) at (18.5, 0.8) {};

\node[font=\sffamily\bfseries] at (18.5, 0.0) {Data Lakehouse};

\fill[kwcolor!5, rounded corners=2pt] (13.3, 0.33) rectangle (23.7, 1.87);

\draw[thick] (13.25, 0.30) -- (23.75, 0.30);

\node[font=\sffamily\bfseries] at (18.5, 1.55) {\textsc{Decisionhouse} Layer};

\node[font=\sffamily\small, align=center] at (14.56, 0.95) {Decision\\[-1pt]Planner};
\node[font=\sffamily\small, align=center] at (17.19, 0.95) {Solver\\[-1pt]Backends};
\node[font=\sffamily\small, align=center] at (19.81, 0.95) {Decision\\[-1pt]Catalog};
\node[font=\sffamily\small, align=center] at (22.44, 0.95) {Caching};

\foreach \x in {15.875, 18.5, 21.125} {
    \draw[gray, thin] (\x, 0.5) -- (\x, 1.4);
}

\node[abox, minimum width=3cm] (desc2) at (15.3, 3.65) {Descriptive Analytics};
\node[abox, minimum width=3cm] (pred2) at (18.5, 3.65) {Predictive Analytics};
\node[abox, minimum width=3cm, line width=1.6pt] (presc2) at (21.7, 3.65)
    {\textbf{Prescriptive Analytics}};

\draw[arr] (stack.north -| desc2) -- (desc2.south);
\draw[arr] (stack.north -| pred2) -- (pred2.south);
\draw[arr, line width=1.4pt] (stack.north -| presc2) -- (presc2.south);

\node[font=\sffamily\small\itshape, fill=white] at (21.7, 2.7)
    {native --- no external pipelines};

\end{tikzpicture}%
}%
\vspace{-4pt}
\caption{Prescriptive analytics in data platforms: per-problem external pipelines today~(a) vs.\ native in a \systemname{}~(b).}
\label{fig:prescriptive-gap}
\end{figure*}
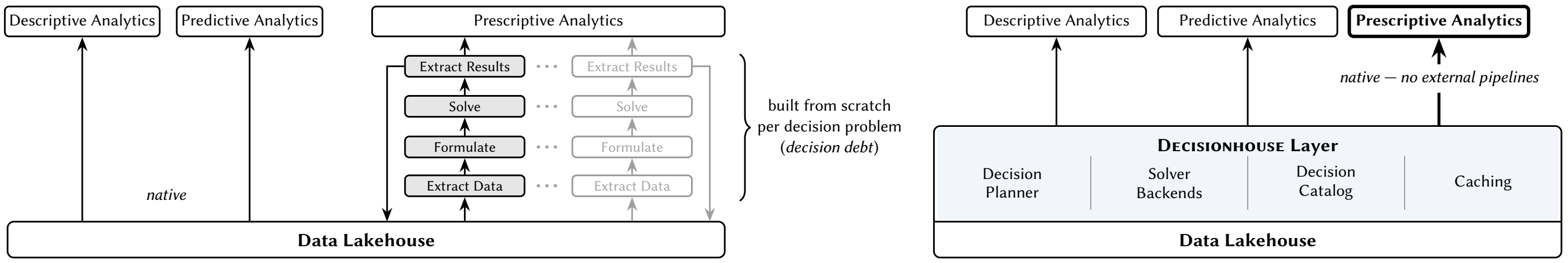

A SQL database can easily answer \emph{``what does it cost to ship from each warehouse to each store?''} with a single join. It cannot answer \emph{``which warehouse-to-store routes should we activate to minimize total cost while meeting every store's demand without exceeding any warehouse's supply?''} This \emph{shipping problem} is a \emph{prescriptive} question~\cite{powell-sda,vldbj-psa-survey,icdt2025-psa}: given options from relational data, constraints from business policy, and a measurable objective, compute the best course of action. \hlRthree{Such questions arise across many domains, including supply chain routing, GPU allocation, shift scheduling, batch sequencing, and portfolio selection.} Answering each today requires leaving the database for a bespoke pipeline that extracts data, encodes a formulation, runs an optimization solver, and maps results back. Each new problem requires its own pipeline. We call this accumulated cost \emph{decision debt} (\Cref{fig:prescriptive-gap}a).

In these pipelines, data extraction, solver invocation, and result mapping have established tooling. Producing the \emph{formulation}, however, is still hard. \hlgen{The formulation specifies the problem's variables, objective, and constraints for a particular \emph{solver} (an algorithm that finds an assignment optimizing the objective subject to the constraints). \emph{Mixed-Integer Linear Programming} (MILP) is one of the most common: continuous or integer variables with linear objective and constraints. MILPs model problems from scheduling and routing to resource allocation and portfolio selection~\cite{wolsey-ip}.}

The shipping problem can be written as a MILP, but not every problem that fits MILP is best solved as one. Shipping itself can equivalently be written as a \emph{minimum-cost flow} problem, provided certain conditions hold: warehouses and stores form a bipartite supply-demand graph, with supply at each warehouse and demand at each store~\cite{ahuja-network-flows}. Conditions like these are called the problem's \emph{structure}. Both formulations produce the same optimum, but the structure determines the complexity: MILP is NP-hard in general, while minimum-cost flow runs in polynomial time. \hlRthree{Many other decision problems exhibit this pattern: resource allocation, worker assignment, and route planning all admit specialized formulations of much lower complexity than generic MILP.}

\hlRfour{Such structure is present in the data and the schema, but generic formulations obscure it:} once committed to MILP, the problem becomes a flat coefficient matrix in which supply caps and demand requirements are indistinguishable rows. \hlRfour{SQL preserves the schema's structure (entities, keys, relationships) but cannot express prescriptive questions. Modeling languages such as AMPL~\cite{ampl}, Pyomo~\cite{pyomo}, and JuMP~\cite{jump} encode a chosen formulation but offer no support for selecting one, and they lack the relational view of the data that makes structure visible.}

What is missing is a layer between data and modeling languages: the \emph{\systemname{}} (\Cref{fig:prescriptive-gap}b), a new class of data infrastructure for prescriptive analytics that internalizes per-problem pipelines behind a declarative interface \hlRfour{that preserves the problem structure}. That interface is \emph{\querylang{}} (Decision Query Language, pronounced ``dee-kwl''), a SQL extension where decision options are declared as relations, each row carries a decision variable, and constraints and objectives are aggregate expressions. \hlRfour{Because these constructs stay anchored to the schema---column names, joins, and group-bys---structural patterns are visible to the system.} The user writes one \querylang{} query per decision problem; the \systemname{} handles data binding, formulation selection, solving, and result mapping. Decision queries produce standard relations, so results compose with SQL and other decision queries without the separate result-mapping step that bespoke pipelines require.

\hlRthree{Given the shipping problem as a \querylang{} query, a \systemname{} can recognize its bipartite supply-demand structure and dispatch the query to a minimum-cost flow solver rather than MILP~(\Cref{sec:architecture}). Shipping is one of many decision problems where bipartite supply-demand structure makes minimum-cost flow applicable. GPU allocation is another such case:} in our experiments, the \querylang{} query runs up to $355\times$ faster than MILP at 1.8M candidate assignments; at 7.4M, MILP times out after 10 minutes while min-cost flow solves in 2.2 seconds. Both formulations produce identical optimal solutions, so the speedup comes entirely from structure recognition (\Cref{sec:deqlite}).

\hlRthree{\systemnameplural{} also give LLMs a target language and runtime for prescriptive analytics: a smaller, verifiable alternative to the free-form solver code generated by today's end-to-end LLM optimization tools~\cite{optiguide,optimus,orlm}. For descriptive analytics, text-to-SQL is mature at the front~\cite{bird-benchmark} and LLM-synthesized engines are emerging at the back~\cite{bespoke-olap,gendb}; \systemnameplural{} open this decomposition to prescriptive analytics: text-to-\querylang{}, \querylang{}-to-code synthesis, and explicit formulation selection in between. End-to-end tools skip this step: the formulation is whatever the LLM emits, often MILP~\cite{llm-meets-optimization,optibench}; each deployment is a standalone script with no specification to verify against. Whether the formulation comes from a rule, a statistic, or an LLM, \querylang{} is the contract it must satisfy.}

This paper defines \textbf{\numrequirements{} requirements} for prescriptive data infrastructure (\Cref{sec:motivation}), presents \textbf{\querylang{}} (\Cref{sec:deql}) and a \textbf{\systemname{} architecture} with a three-IR pipeline and automatic formulation selection (\Cref{sec:architecture}), and outlines \textbf{open problems} at the intersection of database systems, combinatorial optimization, and scalable solving\hlRfour{---including data-centric directions on structure detection, cost models, physical design, responsive execution, and decision governance} (\Cref{sec:research}). This work extends prior research on in-database prescriptive analytics~\cite{paql,spaql,progressive-shading,stochastic-sketchrefine} from single-class package queries to a general-purpose system. The design is informed by our ongoing industry collaborations in logistics, process mining, and cloud platforms.

\section{The Decisionhouse}
\label{sec:motivation}

\looseness=-1
Building a \systemname{} is now feasible and timely. Lakehouse architectures~\cite{lakehouse} and open table formats~\cite{delta-lake} decouple data from its host engine, so any system can read it. PaQL~\cite{paql} absorbed the optimization pipeline for package queries; abstraction-based decomposition~\cite{paql,stochastic-sketchrefine,progressive-shading} scales it to billions of candidates. \hlRthree{Recent LLM capabilities at natural-language-to-optimization translation~\cite{optiguide,optimind,optimus,orlm} demonstrate that this kind of front-end is feasible; in a \systemname{}, the LLM target shifts from solver scripts to \querylang{}, making text-to-\querylang{} a natural LLM front-end on top of the \systemname{}.}

\smallskip\noindent\textbf{Requirements.} \Numrequirements{} requirements together define the \systemname{}.
\reqname{\textbf{(R1) Accessibility}}. Analysts need an interface above integer programming~\cite{optiguide,optimind,optimus,gurobi-ai-assistant}.
\reqname{\textbf{(R2) Data-native}}. Optimization consumes data in place---relational tables, Parquet, Delta Lake---without manual extraction~\cite{solvedb,paql}.
\reqname{\textbf{(R3) Extensibility}}. The system handles diverse patterns (subset selection, resource allocation, assignment, scheduling) and admits new constraint types.
\reqname{\textbf{(R4) Composability}}. The output of one decision query is consumable as input to another, echoing relational algebra's closure property; languages without it exhibit expressibility gaps~\cite{libkin-gql-2025}.
\reqname{\textbf{(R5) Automatic planning}}. Algorithm choice follows from problem structure: a supply-demand query is exponential as MILP but polynomial as network flow.
\reqname{\textbf{(R6) Governance}}. Decisions must be explainable and auditable for regulatory compliance~\cite{fico-platform,aera-decision-cloud}.
\reqname{\textbf{(R7) Guarantees}}. Decisions must be feasible and bounded; solvers already provide optimality gaps and feasibility certificates.
\reqname{\textbf{(R8) Responsiveness}}. Users expect interactive response: progress feedback, time-bounded execution, and best-found solutions within user budgets~\cite{anytime-algorithms,blinkdb}.

\begin{table}[t]
\centering
\footnotesize
\setlength{\tabcolsep}{2pt}
\renewcommand{\arraystretch}{0.85}
\caption{Existing systems vs.\ \systemname{} requirements. \cmarkv~addressed, \pmarkv~partial, \xmarkv~not in scope. Partial notes: \textsuperscript{1}OR expertise required; \textsuperscript{2}data connectors, not integrated; \textsuperscript{3}generic, not extensible; \textsuperscript{4}binary selection only; \textsuperscript{5}logging, not decision-specific; \textsuperscript{6}LLM formulation may be incorrect; \textsuperscript{7}solver-level time limits; \textsuperscript{8}scalable, no progress feedback.}
\label{tab:comparison}
\begin{tabular*}{\columnwidth}{@{}l@{\extracolsep{\fill}}ccccccc@{}}
\toprule
\textbf{Requirement} & \textbf{Solvers} & \shortstack[c]{\textbf{Modeling}\\[-3pt]\textbf{langs}} & \shortstack[c]{\textbf{OR}\\[-1pt]\textbf{Platf.}} & \textbf{PaQL} & \textbf{SolveDB} & \textbf{LLM} & \shortstack[c]{\textbf{\textsc{Decision-}}\\[-1pt]\textbf{\textsc{house}}} \\
\midrule
(R1) Accessible & \xmarkv & \pmarkv\textsuperscript{1} & \pmarkv\textsuperscript{1} & \cmarkv & \pmarkv\textsuperscript{1} & \cmarkv & \cmarkv \\
(R2) Data-native & \xmarkv & \xmarkv & \pmarkv\textsuperscript{2} & \cmarkv & \cmarkv & \xmarkv & \cmarkv \\
(R3) Extensible & \cmarkv & \cmarkv & \cmarkv & \xmarkv & \pmarkv\textsuperscript{3} & \cmarkv & \cmarkv \\
(R4) Composable & \xmarkv & \xmarkv & \xmarkv & \pmarkv\textsuperscript{4} & \cmarkv & \xmarkv & \cmarkv \\
(R5) Auto-plan\vphantom{\textsuperscript{1}} & \xmarkv & \xmarkv & \xmarkv & \xmarkv & \xmarkv & \xmarkv & \cmarkv \\
(R6) Governance & \xmarkv & \xmarkv & \pmarkv\textsuperscript{5} & \pmarkv\textsuperscript{5} & \pmarkv\textsuperscript{5} & \xmarkv & \cmarkv \\
(R7) Guarantees & \cmarkv & \cmarkv & \cmarkv & \cmarkv & \cmarkv & \pmarkv\textsuperscript{6} & \cmarkv \\
(R8) Responsive & \pmarkv\textsuperscript{7} & \pmarkv\textsuperscript{7} & \pmarkv\textsuperscript{7} & \pmarkv\textsuperscript{8} & \xmarkv & \pmarkv\textsuperscript{7} & \cmarkv \\
\bottomrule
\end{tabular*}
\vspace{-4pt}
\end{table}

\looseness=-1
Automatic planning (R5)---the \systemname{}'s defining capability, absent from every other system (\Cref{tab:comparison})---requires detecting structure from the declarative formulation: \hlgen{solvers do limited matrix-level recognition but miss the query-level structure that drives formulation choice}, and LLMs generate solver code end-to-end with no planning stage or correctness guarantees.

\section{\querylang{}: The Decision Query Language}
\label{sec:deql}

\kwSELECT{} answers ``what data exists?'' \querylang{} introduces \kwDECIDE{} to answer ``what actions to take?'' \kwPREDICT{} invokes a pre-trained model; \kwDECIDE{} triggers a translation and optimization pipeline. \querylang{} preserves the relational model end-to-end: queries consume relations and produce relations. We outline the major features below; full details are in the language specification~\cite{deql-language-specs}.

\noindent\textbf{Running example: GPU allocation.}
\hlRfour{An AI inference platform allocates GPU-hours from compute pools to workloads at minimum cost, subject to capacity and demand constraints.}
\Cref{fig:full-example} shows the input tables and query: pools with available capacity, workloads with compute demand, and assignments with costs and latencies.

\begin{figure}[t]
\centering
\footnotesize
\setlength{\fboxsep}{1pt}%
\setlength{\tabcolsep}{1.5pt}%
\begin{minipage}{\linewidth}%
\noindent
\begin{tabular}[t]{>{\columncolor{gray!18}}l>{\columncolor{gray!8}}r}
\multicolumn{2}{c}{\textbf{gpu\_pools}} \\
\hline
pool\_id & capacity \\
\hline
H100-E & 500 \\
A100-W & 300 \\
\end{tabular}%
\hfill
\begin{tabular}[t]{>{\columncolor{gray!18}}l>{\columncolor{gray!8}}r}
\multicolumn{2}{c}{\textbf{workloads}} \\
\hline
workload\_id & demand \\
\hline
llm-serve & 200 \\
embed-batch & 250 \\
fine-tune & 150 \\
\end{tabular}%
\hfill
\begin{tabular}[t]{>{\columncolor{gray!18}}l>{\columncolor{gray!8}}l>{\columncolor{gray!18}}r>{\columncolor{gray!8}}r}
\multicolumn{4}{c}{\textbf{assignments}} \\
\hline
pool\_id & workload\_id & cost & latency\_ms \\
\hline
H100-E & llm-serve & 10 & 45 \\
H100-E & embed-batch & 15 & 120 \\
A100-W & embed-batch & 8 & 85 \\
A100-W & fine-tune & 14 & 160 \\
\end{tabular}%
\end{minipage}
\vspace{0pt}
\ifarxiv\vspace{4pt}\fi
\begin{lstlisting}[style=deql]
CREATE CANDIDATES gpu_assignments
DECISION KEY (pool_id, workload_id) AS
  SELECT a.pool_id, a.workload_id,
         p.capacity, w.demand, a.cost, a.latency_ms
  FROM assignments a
  JOIN gpu_pools p ON a.pool_id = p.pool_id
  JOIN workloads w ON a.workload_id = w.workload_id;

DECIDE allocation_plan
FROM gpu_assignments
DECISION COLUMNS (
    active   SELECTION BINARY,
    quantity CONTINUOUS BETWEEN 0 AND capacity)
WHERE latency_ms <= 200
SUBJECT TO
  CONSTRAINT pool_capacity: SUM(quantity) <= capacity BY pool_id,
  CONSTRAINT meet_demand:   SUM(quantity) = demand BY workload_id
MINIMIZE SUM(cost * quantity);
\end{lstlisting}
\ifarxiv\vspace{-2pt}\fi
\caption{GPU allocation example: input tables (top) and \querylang{} workflow (bottom).}
\label{fig:full-example}
\ifarxiv\vspace{12pt}\fi
\end{figure}

\kwCREATECANDIDATES{} factors the candidate definition into a first-class database construct: a named query like a view, annotated with a \kwDECISIONKEY{} that declares which columns identify each \emph{decision entity} (metadata that neither tables nor views carry). In the GPU example, \hlRfour{the \kwSELECT{} joins three tables to produce one row per pool--workload pair; each such row is a \emph{candidate}, and} \kwDECISIONKEY{}~\texttt{(pool\_id, workload\_id)} declares that each pair is one decision entity.

The \kwDECIDE{} statement optimizes over the candidates and materializes the result as a new table (e.g., \texttt{allocation\_plan}). \kwDECISIONCOLUMNS{} declares the \emph{decision variables} (one per candidate): unknowns computed at query time, unlike data columns which are read from the database; they otherwise behave like regular columns in expressions and output. Each variable declares a domain: \kw{BINARY} for select/not-select, or \kw{CONTINUOUS} with data-driven bounds (e.g., \texttt{BETWEEN 0 AND capacity}). A \kwBY{} clause instantiates one constraint per group: \hlRfour{with two pools (H100-E, A100-W), \texttt{SUM(quantity) <= capacity BY pool\_id} generates two constraints, one capping each pool's allocation.} The optional \kw{SELECTION} modifier makes \texttt{active} a binary on/off switch: when \texttt{active}${}=0$, \texttt{quantity} is forced to zero and the row is omitted from the result, so this linking need not be written by hand (generalizing PaQL's~\cite{paql} package semantics). The \kw{MINIMIZE}/\allowbreak\kw{MAXIMIZE} clause specifies the objective; appending \texttt{WITHIN X\%} bounds the optimality gap, and appending \texttt{TIMEOUT X s} caps solve time, letting users trade optimality for speed~(\Cref{sec:dqop}).

Three further constructs complete the sketch. \kw{WHERE} in \kwCREATECANDIDATES{} removes physically infeasible candidates; \kw{WHERE} in \kwDECIDE{} applies per-query policy filters (e.g., latency SLAs). Multi-level problems (assigning items to batches while choosing each batch's timing) join candidate tables in \kwDECIDE{}: each keeps its own \kwDECISIONKEY{}, so its decision variables are not duplicated across the join. Stochastic extensions attach distributional annotations to data columns and probabilistic guarantees to constraints. The language specification~\cite{deql-language-specs} covers each in detail.

\kwCREATEABSTRACTION{} defines a coarser view of the candidate space---e.g., clustering millions of GPU assignments into thousands of representative groups---enabling decomposition methods~\cite{aggregation-disaggregation} like SketchRefine~\cite{paql} and partial-state abstraction~\cite{mdpabstraction} to meet \texttt{WITHIN}/\texttt{TIMEOUT} budgets at scale:
\vspace{1pt}
\needspace{4\baselineskip}
\begin{lstlisting}[style=deql]
CREATE ABSTRACTION assignment_clusters
ON gpu_assignments (cost, demand, capacity) USING KMEANS(k=1000);
\end{lstlisting}
\vspace{1pt}\noindent SketchRefine solves a reduced problem on representatives, then refines only promising clusters, shrinking effective problem size by orders of magnitude within bounded error. We discuss extensions to other problem classes and anytime variants in \Cref{sec:research}.

\hlRfour{\querylang{} rests on relational algebra: \kwCREATECANDIDATES{} takes a SQL query as body, grouped constraints use \kw{GROUP BY} semantics, and column references are relational attributes. Extending SQL reuses these primitives directly. Beyond syntax, the relational grounding yields several benefits: \kwDECIDE{} produces a relation, so decision queries compose with SQL and with other \kwDECIDE{} queries without an export/import step; the decision planner reads candidate-set statistics for formulation selection (\Cref{sec:dqop}); and candidate sets and decision outputs reuse the host store's physical design and governance (provenance, audit, replay). This mirrors in-database ML extensions like BigQuery ML and Snowflake ML, which add new declarative constructs (e.g., \kw{CREATE MODEL}) to SQL while pure-SQL operations work unchanged.} \hlRthree{The same primitives already express different optimization problems---subset selection, bipartite matching, multi-level batching, and job-shop scheduling (worked through in the language specification~\cite{deql-language-specs})---and like SQL, \querylang{} is designed to grow with new constructs and keywords over time (\Cref{sec:research}).}

\section{Decision Query Processing}
\label{sec:architecture}

\definecolor{nodepoolcolor}{RGB}{40,130,130}      %
\definecolor{nodeworkloadcolor}{RGB}{205,115,90}  %

\begin{figure}[t]
\centering
\setlength{\abovecaptionskip}{2pt}
\resizebox{\columnwidth}{!}{%
\begin{tikzpicture}[
  >=Stealth,
  outer sep=0pt,
  irlabel/.style={font=\bfseries\sffamily, text=black, anchor=base},
  exbox/.style={anchor=center},
]

\def\colA{0}
\def\colB{3.0}
\def\colC{5.7}

\node[irlabel] (lbl_a) at (\colA-0.35, 0.85) {Abstract IR};
\node[irlabel] (lbl_b) at (\colB, 0.85) {Concrete IR};
\node[irlabel] (lbl_c) at (\colC, 0.85) {Materialized IR};

\draw[->, thick, black] ([xshift=3pt]lbl_a.east) -- ([xshift=-3pt]lbl_b.west |- lbl_a.east);
\draw[->, thick, black] ([xshift=3pt]lbl_b.east) -- ([xshift=-3pt]lbl_c.west |- lbl_b.east);

\node[font=\scriptsize\sffamily, text=black, align=left, anchor=center, inner xsep=0pt] (ex_a) at (\colA-0.05, 0.17)
  {\textbf{\textsc{decision key}}: (pool\_id, workload\_id)\\[-0.5pt]
   \textbf{\textsc{decision cols}}: active, quantity\\[-0.5pt]
   \textbf{\textsc{filters}}: latency\_ms $\le$ 200\\[-0.5pt]
   \textbf{\textsc{constraints}}: pool\_capacity, meet\_demand\\[-0.5pt]
   \textbf{\textsc{linear objective}}: min $\Sigma$ cost $\cdot$ qty};

\begin{scope}[shift={(\colB, 0.05)}]
  \fill[nodepoolcolor] (-0.6, 0.25) circle (3pt);
  \fill[nodepoolcolor] (-0.6, -0.25) circle (3pt);
  \draw[nodeworkloadcolor, fill=nodeworkloadcolor, thick] ([xshift=-2.5pt, yshift=-2.5pt]0.6, 0.35) rectangle ++(5pt, 5pt);
  \draw[nodeworkloadcolor, fill=nodeworkloadcolor, thick] ([xshift=-2.5pt, yshift=-2.5pt]0.6, 0) rectangle ++(5pt, 5pt);
  \draw[nodeworkloadcolor, fill=nodeworkloadcolor, thick] ([xshift=-2.5pt, yshift=-2.5pt]0.6, -0.35) rectangle ++(5pt, 5pt);
  \draw[black!40, thin] (-0.6, 0.25) -- (0.6, 0.35);
  \draw[black!40, thin] (-0.6, 0.25) -- (0.6, 0);
  \draw[black!40, thin] (-0.6, -0.25) -- (0.6, 0);
  \draw[black!40, thin] (-0.6, -0.25) -- (0.6, -0.35);
  \node[font=\scriptsize, text=nodepoolcolor, anchor=base] at (-0.6, 0.55) {\texttt{pool\_id}};
  \node[font=\scriptsize, text=nodeworkloadcolor, anchor=base] at (0.6, 0.55) {\texttt{workload\_id}};
\end{scope}

\begin{scope}[shift={(\colC, 0.05)}]
  \fill[nodepoolcolor] (-0.6, 0.25) circle (3pt);
  \fill[nodepoolcolor] (-0.6, -0.25) circle (3pt);
  \draw[nodeworkloadcolor, fill=nodeworkloadcolor, thick] ([xshift=-2.5pt, yshift=-2.5pt]0.6, 0.35) rectangle ++(5pt, 5pt);
  \draw[nodeworkloadcolor, fill=nodeworkloadcolor, thick] ([xshift=-2.5pt, yshift=-2.5pt]0.6, 0) rectangle ++(5pt, 5pt);
  \draw[nodeworkloadcolor, fill=nodeworkloadcolor, thick] ([xshift=-2.5pt, yshift=-2.5pt]0.6, -0.35) rectangle ++(5pt, 5pt);
  \draw[black!50, thin] (-0.6, 0.25) -- (0.6, 0.35) node[midway, above=-2pt, font=\scriptsize, text=black] {10};
  \draw[black!50, thin] (-0.6, 0.25) -- (0.6, 0) node[midway, above=-2pt, font=\scriptsize, text=black] {15};
  \draw[black!50, thin] (-0.6, -0.25) -- (0.6, 0) node[midway, below=-2.6pt, font=\scriptsize, text=black] {8};
  \draw[black!50, thin] (-0.6, -0.25) -- (0.6, -0.35) node[midway, below=-2pt, font=\scriptsize, text=black] {14};
  \node[font=\scriptsize, text=black, anchor=east] at (-0.7, 0.25) {500};
  \node[font=\scriptsize, text=black, anchor=east] at (-0.7, -0.25) {300};
  \node[font=\scriptsize, text=black, anchor=west, inner xsep=0pt] at (0.76,0.35) {200};
  \node[font=\scriptsize, text=black, anchor=west, inner xsep=0pt] at (0.76,0) {250};
  \node[font=\scriptsize, text=black, anchor=west, inner xsep=0pt] at (0.76,-0.35) {150};
  \node[font=\scriptsize, text=nodepoolcolor, anchor=base] at (-0.6, 0.55) {\texttt{pool\_id}};
  \node[font=\scriptsize, text=nodeworkloadcolor, anchor=base, inner xsep=0pt] at (0.6, 0.55) {\texttt{workload\_id}};
\end{scope}

\end{tikzpicture}%
}%
\caption{\hlRfour{GPU query at each IR: Abstract (symbolic), Concrete (min-cost flow, symbolic), Materialized (data-populated).}}
\label{fig:ir-walkthrough}
\end{figure}
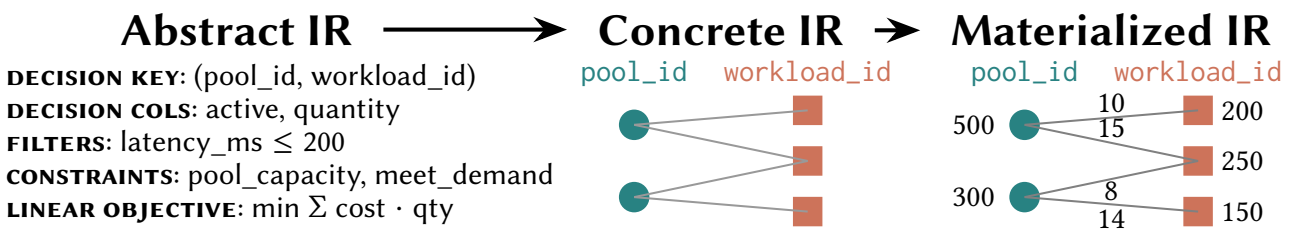

\hlRthree{A \systemname{} ingests \querylang{} queries and returns decisions as new relational tables. The query-processing pipeline comprises four components connected by three intermediate representations, each marking a distinct commitment (\Cref{fig:ir-walkthrough} traces the GPU query through them). The \textbf{Parser} produces \textbf{Abstract IR}: the decision problem stated symbolically over schema columns---variables, grouped constraints, objective. The \textbf{Decision Planner} produces \textbf{Concrete IR}, the chosen formulation (still symbolic), by detecting structural patterns in the Abstract IR~(\Cref{sec:dqop}): for the GPU query, bipartite supply-demand structure selects min-cost flow over generic MILP. Committing the formulation against database statistics before materialization lets one Concrete IR be reused across many Materialized IRs, one per data subset (SketchRefine-style decomposition~\cite{paql}) or per parameter setting (what-if analysis).}

\hlRtwo{The \textbf{Materializer} produces \textbf{Materialized IR} by binding Concrete IR's symbolic expressions to data fetched from federated sources (e.g., lakehouse, operational database, ML pipeline). The \textbf{Executor} dispatches Materialized IR to the solver (e.g., HiGHS\slash Gurobi\slash SCIP for LP/MILP, OR-Tools' cost-scaling solver for min-cost flow), writes the solution to the \kwDECIDE{} output table, addressable by later SQL or \querylang{} queries, and logs the decision context (query, formulation, solver parameters, solution, timestamp) for audit and replay. This decoupling of formulation choice from solver dispatch lets future backends (e.g., GPU-accelerated LP solvers, network-flow hardware) plug in without changing the planner.}

\subsection{The Three-IR Design}
\label{sec:three-ir}

The Abstract-to-Concrete boundary plays the same role as SQL planning: it separates \emph{what} to optimize from \emph{how} to formulate it. Because Abstract IR is symbolic, type checking, name resolution, and linearity checks all happen against it before data is fetched, rejecting invalid queries (e.g., non-linear objectives) early. The Concrete-to-Materialized boundary is what optimization adds beyond SQL planning: solving requires a global view of the data---a single numeric instance assembled from candidate rows---which must be materialized as a distinct stage. The materialization strategy depends on the formulation: a network flow needs supply and demand vectors indexed by node, while a generic MILP needs a flat coefficient matrix. Deferring data binding until planning is finalized lets the planner choose both the formulation and its materialization strategy~\cite{composable-manifesto,boss}.

\textbf{Abstract IR} is the parser's structured representation of the problem. For the GPU query (\Cref{fig:full-example}), it contains two decision variables keyed on \texttt{(pool\_id, workload\_id)}---a binary \texttt{active} flag and a continuous \texttt{quantity}---supply and demand constraints grouped via \kwBY{}, and a linear cost-minimization objective, all stated symbolically over column references. Shortcuts like \kwSELECTION{} desugar to explicit constraints (details in the language specification~\cite{deql-language-specs}).

\looseness=-1
\hlRfour{\textbf{Concrete IR} encodes the chosen formulation. Growing the family of targets across new problem classes (and the structure-detection logic that recognizes them) is part of the research agenda (\Cref{sec:research}). Current targets include a generic MILP with a general constraint matrix, an LP relaxation that drops integrality, and a min-cost network flow as a bipartite supply-demand graph. For the GPU query, the planner selects min-cost flow because the bipartite \kwDECISIONKEY{} and supply-demand \kwBY{} constraints match flow's structure (\Cref{sec:dqop} details the criteria); the middle column of \Cref{fig:ir-walkthrough} shows the resulting graph template.}

\textbf{Materialized IR} is the data-populated form: the Concrete IR's symbolic expressions evaluated against actual rows. For the GPU query, the right column of \Cref{fig:ir-walkthrough} populates the bipartite graph directly from \Cref{fig:full-example}'s three tables: pool nodes carry capacities from \texttt{gpu\_pools}, workload nodes carry demands from \texttt{workloads}, and arcs carry costs from \texttt{assignments}---vectors that a min-cost-flow solver consumes directly. For LP/MILP targets the Materialized IR is the flat coefficient form (objective vector $\mathbf{c}$, constraint matrix $\mathbf{A}$, bound vectors $\mathbf{l},\mathbf{u}$) instead of a graph. Decomposition methods like SketchRefine~\cite{paql,progressive-shading,stochastic-sketchrefine} reuse one Concrete IR across many Materialized IRs---once over cluster representatives, then once per refined cluster.

\vspace{2pt}
\noindent\hlRtwo{\textbf{Correctness.} Each stage of the three-IR pipeline (Abstract IR, Concrete IR, Materialized IR) must preserve the original DeQL query's constraints and objective. To establish this, it suffices to verify each transformation independently---desugaring, structure detection, formulation choice, materialization. Decomposition methods carry bounded-error theorems (e.g., PaQL/sPaQL~\cite{paql,spaql} for subset \kwSELECTION{}). \kw{WITHIN}/\kw{TIMEOUT} budgets bound the solver's output. Extending the algebraic foundation~\cite{pratten-subset-algebra} across the full language (including stochastic and relaxation~\cite{package-relaxation} constructs) is an important research direction~(\Cref{sec:research}).}

\subsection{The Decision Planner}
\label{sec:dqop}

MILP is the planner's default target, and structural patterns in the query enable more efficient polynomial-time alternatives such as min-cost flow~\cite{ahuja-network-flows}, LP relaxation, or the Hungarian algorithm~\cite{kuhn-hungarian} when they apply. This choice can change the effective complexity class from NP-hard to polynomial (\Cref{sec:deqlite}). We call the problem of selecting the fastest applicable formulation from a declarative query the \emph{Decision Optimization Problem} (\optimizername{}). Solvers optimize \emph{within} a fixed coefficient matrix; the decision planner optimizes \emph{which instance} the solver receives, using query structure and database statistics that no solver has access to.

\vspace{2pt}
\noindent\hlRtwo{\textbf{Planner objective.} The planner minimizes solve time subject to the user's quality budget (\kw{WITHIN}~$X$\% or \kw{TIMEOUT}; optimal by default). A simple rule yields large gains: prefer any applicable polynomial-time formulation (min-cost flow, LP relaxation, Hungarian) over MILP when structure detection succeeds, since the complexity-class ordering is known in advance and the gap is often orders of magnitude (\Cref{sec:deqlite}). Refining this with a proper cost model (particularly for ranking multiple polynomial alternatives when several apply) is central to our research agenda~(\Cref{sec:research}).}

\noindent\textbf{Structure detection from Abstract IR.} The Abstract IR preserves structural patterns that flat coefficient matrices obscure. In the GPU allocation query (\Cref{fig:full-example}), the \kwDECISIONKEY{} \texttt{(pool\_id, workload\_id)} reveals bipartite structure; the \kwBY{} constraints match supply-demand flow conservation; the objective is linear in the flow variable. Together, these identify a min-cost network flow, enabling min-cost flow instead of branch-and-bound (\Cref{sec:deqlite}). Single-component constraint groups with bare-variable sums reveal total unimodularity, so LP relaxation is exact (Hoffman--Kruskal~\cite{ahuja-network-flows}); bipartite queries whose constraint bounds are the literal one (e.g., \texttt{COUNT(*) = 1}) reduce to assignment, solvable by the Hungarian algorithm~\cite{kuhn-hungarian} in $O(n^3)$. Each is a deterministic pattern match on the IR, analogous to recognizing star joins from query graphs. \hlRthree{The same patterns recur across decision workloads: bipartite supply-demand structures transportation, ad placement, and resource-allocation problems; total unimodularity holds for bipartite matching and assignment~\cite{ahuja-network-flows}. Extending the planner to detect further patterns is part of the research agenda (\Cref{sec:research}).}

\noindent\textbf{Statistics-driven planning.}
\looseness=-1
Not all structure is visible in the IR; some shows up only in the candidate data. The planner reads per-column statistics on the candidate set (\kw{MIN}, \kw{MAX}, \kw{COUNT DISTINCT}, all-integer flag) computed at \kwCREATECANDIDATES{} time as a byproduct of the data scan, analogous to SQL \kw{ANALYZE}. \texttt{COUNT(DISTINCT cost)} = 1 on a min-cost flow query with equality demands reduces it to max-flow, since a constant cost objective then makes any feasible flow optimal. Coefficients confined to $\{-1, 0, 1\}$ on candidate columns can complete a data-dependent total-unimodularity check~\cite{ahuja-network-flows} that the IR alone cannot finish, reducing MILP to LP relaxation~(\Cref{sec:research}).

\noindent\textbf{Planning under uncertainty.} Stochastic queries~\cite{spaql,stochastic-sketchrefine} introduce uncertainty in input attributes, adding another dimension to planning~(\Cref{sec:research}).

\subsection{Analysis}
\label{sec:deqlite}

We are building \productname{}, an initial \systemname{} implementation in Rust, using Apache DataFusion~\cite{datafusion} for query federation, Apache Arrow~\cite{arrow} as columnar format, and HiGHS~\cite{highs}/Gurobi~\cite{gurobi} for solving. The current release includes a \querylang{} parser, all three IRs, a working SketchRefine implementation, and direct solving. Data sources include DuckDB, ODBC databases, Microsoft Fabric, and all DataFusion-supported backends. \querylang{} Studio, an interactive interface to this prototype, is demonstrated in a companion paper~\cite{deql-studio-demo}.

\begin{figure}[t]
\centering
\includegraphics[width=\columnwidth]{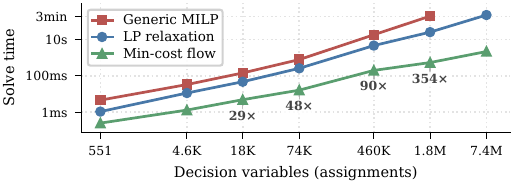}
\caption{Solve time for the GPU allocation query (\Cref{fig:full-example}). All three formulations produce identical optima up to 1.8M assignments; min-cost flow is 29--355$\times$ faster than MILP. At 7.4M, MILP times out (10~min); min-cost flow solves in 2.2~s.}
\label{fig:formulation-scaling}
\end{figure}

\noindent\textbf{Formulation selection.}
\looseness=-1
To evaluate automatic structure recognition, we solve the GPU allocation query from \Cref{fig:full-example} at seven scales up to 7.35M candidate assignments (consistent with hyperscale GPU clusters that reach $O(10^5)$ GPUs~\cite{tang-meta-hyperscale,kumar-hyperscale-osdi24}) using three formulations that the decision planner would select between: generic MILP (no recognition), LP relaxation (TU detection drops binary variables), and min-cost network flow (full network structure detection). Generic MILP and LP relaxation run on HiGHS~\cite{highs} (branch-and-bound and dual simplex, respectively); min-cost flow runs on OR-Tools'~\cite{ortools} cost-scaling solver. Each formulation has a 10-minute time limit on random instances (median of 3 runs; \Cref{fig:formulation-scaling}).

Speedup grows superlinearly (29$\times$ at 18K, 90$\times$ at 460K, 355$\times$ at 1.8M---from over 3 minutes to under 1 second). All three formulations return provably optimal solutions with identical objective values where they complete; the speedup comes entirely from structure recognition. The transportation matrix is totally unimodular~\cite{ahuja-network-flows}, so LP relaxation produces integer solutions automatically (binary variables redundant); recognizing bipartite supply-demand additionally enables min-cost flow, changing the complexity class. At scale, structure recognition determines whether the query succeeds at all; a declarative system can detect and exploit structure the user did not specify. At 7.4M, MILP fails to find any feasible solution in 10 minutes while min-cost flow returns the optimum in 2.2 seconds. \hlRthree{Which formulations apply depends on the query: without bipartite structure, min-cost flow is unavailable, and LP relaxation or MILP applies depending on total unimodularity.} This experiment isolates one dimension of the \optimizername{}: structure detection from the Abstract IR. Statistics-driven detection from candidate data is a second dimension (\Cref{sec:dqop}); execution-strategy choices like decomposition and pruning are a third dimension, an open research direction~(\Cref{sec:research}).

\noindent\textbf{Incremental materialization.} When successive \kwDECIDE{} queries share the same Abstract and Concrete IR---the common case in what-if analysis, where an analyst varies constraint parameters while the formulation stays fixed---only the Materialized IR changes, and only partially: bound vectors update but the constraint matrix sparsity pattern is unchanged. The system retains both the Materialized IR and the solver's internal state (preprocessed model and factored simplex basis), propagating only the changed bounds through the solver's incremental update API. The solver re-solves from its previous basis without reconstructing the model. On the GPU allocation benchmark, varying pool capacities across 50~what-if queries yields a per-query speedup that grows with problem size once the model is built and reused: 8--11$\times$ at 460K~assignments and 12--15$\times$ at 1.8M.

\section{Research Agenda}
\label{sec:research}

We outline eight research directions tightly coupled to the \systemname{} architecture.

\noindent\hlRtwo{\textbf{Formal foundations for \querylang{}.} \querylang{} needs an algebraic foundation analogous to relational algebra for SQL, covering typed decision variables, multi-table candidate spaces, grouped constraints, and non-linear or stochastic constructs (quadratic objectives, constraint programming, inline ML-model scoring via \kwPREDICT{}, probabilistic predicates). Prior work covers subset selection~\cite{paql,spaql} and a relational algebra for decision queries via exponentiation~\cite{pratten-subset-algebra}; extending to stochastic, probabilistic, and ML-augmented constructs remains open. Each new construct is admissible only via a translation rule and an equivalence theorem, making \querylang{} extensible.}

\noindent\textbf{Automatic reformulation and structure detection.}
\looseness=-1
\hlgen{Current systems have limited support for structure detection: CPLEX's network extractor~\cite{cplex} recovers pure node-arc substructure with specialized heuristics~\cite{brown-wright-1984}, but broader semantic patterns are lost once the problem is flattened to a coefficient matrix. The Abstract IR preserves these patterns. Detecting them draws on database query plan pattern matching~\cite{selinger-access-path} and OR reformulation theory~\cite{savelsbergh-presolve,conforti-extended-formulations,sherali-adams-rlt}.} \hlRthree{New pattern detectors should plug into the planner without modifying the rest of the pipeline.} \hlRthree{Where syntactic rules and statistical methods fall short, the planner can draw on LLM judgment as a third input, with the \querylang{} query serving as the contract any chosen formulation must satisfy.}

\noindent\textbf{Cost models for decision query optimization.} \optimizername{} requires predicting execution cost and solution quality before committing to a plan. MILP solve time resists closed-form characterization; prediction errors of 10--100$\times$ are common even with ML models trained on millions of instances~\cite{solver-runtime-prediction}, partly because predictions operate on flat coefficient matrices with no semantic information. The three-IR design exposes a richer feature space: problem type (assignment vs.\ scheduling), constraint topology (bipartite keys, grouped supply-demand), candidate set size per dimension, and abstraction depth---analogous to the cardinality estimates and index statistics that make database cost models work.

\noindent\textbf{Physical design for decision queries.} Candidate sets introduce physical design questions absent from traditional query processing. Whether a candidate set should be virtual, in-memory, or persisted depends on join complexity, reuse across queries, solver access patterns, and data freshness. Physical layout must serve solver access patterns, which differ from query scans: the join structure in \kwCREATECANDIDATES{} groups candidates by dimension key, defining constraint matrix sparsity patterns to precompute once and reuse. When candidate sets derive from large joins, full materialization may be infeasible; factorized representations~\cite{factorized-databases} avoid materializing joins for aggregation, but computing solver coefficients directly from factorized data remains open. The same materialization question applies to \kwDECIDE{} output, which appends new decision columns to existing data columns.

\noindent\textbf{Responsive execution.} Decisions are coupled through shared constraints, so the row- and batch-independent scaling SQL and ML rely on does not apply directly. Decomposition methods like SketchRefine~\cite{paql} partition the candidate space into subproblems and coordinate at the boundaries. The three-IR pipeline can then compute bounds at coarse abstraction levels before full materialization, and fall back to a different Concrete IR when the solver stalls. This mirrors adaptive query processing~\cite{eddies}, except that adaptation crosses formulations rather than access paths. Several questions remain open. First, abstraction-based decomposition must be extended beyond package queries to scheduling, network flow, and assignment. Second, abstractions need to be designed so that they yield valid solutions at any interruption point. \hlRtwo{Third, scheduling across heterogeneous solver backends (where the formulation/dispatch split admits parallel formulation racing and per-block routing) requires new cost models and routing policies.}

\noindent\hlRthree{\textbf{Optimization-workload benchmarks.} The community needs a TPC-H equivalent for prescriptive workloads---spanning data access and optimization: diverse problem types (knapsack, assignment, scheduling), realistic scales, multiple constraint patterns, and ground-truth optima to compare systems and measure progress.}

\noindent\hlRthree{\textbf{Decision governance.} Decisions that affect people---hiring, allocation, scheduling---require an audit trail, infeasibility explanations, and traceability of outputs through formulation choices. The GDPR's ``right to explanation'' applies to automated decisions, yet no infrastructure directly supports it.}

\noindent\hlRthree{\textbf{LLM-driven prescriptive pipelines.} The decomposition databases are adopting (text-to-SQL~\cite{bird-benchmark,spider2} + SQL-to-code synthesis~\cite{bespoke-olap,gendb}) maps onto prescriptive analytics as text-to-\querylang{} + \querylang{}-to-code synthesis, with the declarative query as the stable intermediate. Text-to-\querylang{} benchmarks, validation methods for generated queries (a missing constraint silently yields an optimal solution to the wrong problem), and LLM-synthesized specialized execution are open problems.}

\section{Related Work}
\label{sec:related}

\looseness=-1
\noindent\textbf{Optimization tools and platforms.} Commercial (Gurobi~\cite{gurobi}, CPLEX~\cite{cplex}\hlnew{, FICO Xpress Solver~\cite{fico-xpress-solver}}) and open-source (HiGHS~\cite{highs}, SCIP~\cite{scip}) solvers provide the algorithmic foundation for optimization. Modeling languages (AMPL~\cite{ampl}, Pyomo~\cite{pyomo}, JuMP~\cite{jump}, MiniZinc~\cite{minizinc}, OR-Tools) add expressivity but require optimization expertise and lack database integration. OR platforms (\hlnew{FICO Xpress Insight~\cite{fico-xpress-insight} and DecisionOps~\cite{fico-decisionops}}, AIMMS~\cite{aimms}, IBM CPLEX Studio~\cite{cplex}) bundle these with data connectors and enterprise logging, covering everything from formulation onward but leaving formulation manual: models come in platform-specific languages, data is mapped imperatively, and formulations are selected by hand. Lakehouse integration runs through notebooks~\cite{gurobi-databricks}. Algorithm selectors such as AutoFolio~\cite{autofolio} choose among solvers from instance features but operate on already-serialized coefficient matrices; they cannot detect that a MILP encodes network-flow or transportation structure. Supply chain platforms~\cite{kinaxis-maestro,o9-solutions} embed domain-specific heuristics for planning but are not general-purpose. A \systemname{} sits above all of these, using solvers as backends while providing data-native interfaces and automatic planning.

\noindent\textbf{LLMs and optimization.} LLMs that translate natural-language descriptions into solver code~\cite{optiguide,optimind,optimus,optimus03,orlm,gurobi-ai-assistant} provide what we call \emph{operational intelligence} (understanding which situation to model). They do not provide \emph{mathematical intelligence}: each generated formulation is a one-off solver script with no specification to verify against, accuracy degrades with combinatorial structure and size~\cite{optibench,llm-meets-optimization,orqa}, and no shared planning layer transfers across queries. When frontier models attempt to produce decisions directly without a solver, feasibility is unreliable across both models and domains~\cite{constraintbench}. We envision LLMs as complementary front-ends, translating user intent into \querylang{}, while a \systemname{} provides verified translation, formulation selection, and guaranteed solving.

\noindent\textbf{Decision support systems.} Classical DSS~\cite{sprague-dss} provide dashboards, what-if analysis, and scenario comparison to aid human judgment. A \systemname{} computes optimal decisions from declarative specifications; classical DSS presents alternatives for manual evaluation.

\noindent\textbf{In-database optimization.} An early SQL-based decision query language was proposed by Brodsky et al.~\cite{brodsky-dql}. SolveDB~\cite{solvedb} embedded solvers in PostgreSQL via UDFs; User-Defined Operators~\cite{udo} compile custom algorithms into the query pipeline with zero-copy data access, but in both cases the solver executes after formulation selection is fixed; the database planner cannot see the mathematical structure. PaQL~\cite{paql} pioneered declarative package queries, scaled by SketchRefine~\cite{aggregation-disaggregation} to millions of candidates and Progressive Shading~\cite{progressive-shading} to billions; Pratten et al.~\cite{pratten-subset-algebra} provide an algebraic foundation for subset selection and optimization in relational algebra; DeciDB~\cite{decidb} extends DuckDB with a \kwDECIDE{} clause inside \kwSELECT{}. \querylang{} makes \kwDECIDE{} a top-level statement and adds a structure-aware planner that detects properties of the constraint matrix (total unimodularity, network-flow shape, block-angular decomposability) and selects among LP, MILP, and decomposition formulations, which none of these systems, including commercial ones like LogicBlox~\cite{logicblox} and RelationalAI~\cite{rel}, detect. \querylang{} extends SQL rather than replacing it, unlike SaneQL~\cite{saneql}.

\section{Conclusion}
\label{sec:conclusion}

SQL and query optimizers eliminated bespoke data-retrieval programs; Lakehouses added first-class support for ML. Deciding (computing optimal actions under constraints) is next. As autonomous systems take on consequential decisions over data, the decision debt of ad-hoc pipelines---hard to audit, reproduce, or govern---is not sustainable. \systemnameplural{} provide the missing layer: \querylang{} decisions are declarative, automatically planned across formulations and solvers, and auditable end-to-end.

\ifextended\else\balance\fi
\bibliographystyle{ACM-Reference-Format}
\bibliography{references}

\end{document}